\documentstyle[preprint,aps,epsf,floats]{revtex}
\tighten

\begin{document}

\preprint{\vbox{
\hbox{CMU-HEP96-15}
\hbox{DOE/ER/40682-126}
\hbox{DOE/ER/41014-06-N97}
}}
\title{Heavy Baryon Mixing in Chiral Perturbation Theory}
\author{C. Glenn Boyd\footnote{{\tt boyd@fermi.phys.cmu.edu}}
and
Ming Lu\footnote{{\tt lu@fermi.phys.cmu.edu}}
}
\address{
Department of Physics, Carnegie Mellon University, 
\\
Pittsburgh, PA 15213
}
\author{Martin  J. Savage\footnote{{\tt 
savage@thepub.phys.washington.edu}}
}
\address{
Department of Physics, University of Washington, 
\\
Seattle, WA 
98195
}
\maketitle

\begin{abstract}

We discuss the SU(3) and heavy quark spin-symmetry breaking
mixing between the $\Xi_c$ and $\Xi_c^\prime$ 
charmed baryons. 
Chromomagnetic hyperfine interactions are the leading source of 
spin-symmetry breaking and together with the SU(3) breaking 
mass differences between the
lightest pseudo-Goldstone bosons gives the leading contribution 
to the mixing.  Such contributions 
are computed in chiral perturbation theory and compared
to quark model expectations. 
We also compute the leading contribution to the semileptonic decay
$\Xi_b\rightarrow \Xi_c^\prime l\overline{\nu}_l$ at zero recoil,
and find that it is an order of 
magnitude smaller than naive power counting would suggest.
It appears that
$\Xi_b\rightarrow \Xi_c^\prime l\overline{\nu}_l$
is dominated by incalculable counterterms, and we discuss the
implications for quark models based on the essential role of
hyperfine interactions.

\end{abstract}

\bigskip
\vskip 2.0cm
\leftline{December 1996}

\vfill\eject

The ongoing experimental efforts in charmed baryon 
physics have established the masses and
dominant decay modes of most of the lowest lying 
charmed baryons.
The  
$\Lambda_c^+$, $\Xi_c^{+,0}$, $\Sigma_c^{++, +, 0}$, and the 
$\Omega_c$ baryons with  $J^\pi={1\over 2}^+$ have 
been observed and studied\cite{expts,delsigb}~,
while just recently the $J^\pi={3\over 2}^+ \,$ 
$\Sigma_c^*$ and $\Xi_c^{*}$ have been observed
\cite{cleoxi,CLEOb}
(we note that a $J^\pi={1\over 2}^+$ $\Xi_c^\prime$ 
baryon candidate has 
also been identified\cite{WAa}\ 
but needs to be verified).
The observed hyperfine mass splitting between the $\Sigma_c^*$ 
and the $\Sigma_c$ of
$\delta_c\sim 67 {\rm MeV}$ is consistent with the 
naive estimate of $\Lambda_{\rm QCD}^2/m_c$ from heavy quark 
considerations, but 
the $\Sigma^*_b - \Sigma_b$ mass splitting that has recently been hinted at
\cite{delsigb}\ 
seems to be larger than heavy quark symmetry predicts.  
To address this issue, 
Falk\cite{falkBary} has proposed a reassignment of
the observed resonances that would maintain the expected
mass splittings. We will 
assume $\delta_b\sim \delta_c/3$, consistent with heavy quark
symmetry predictions.

Despite the mass of the charm quark being only 
$m_c\sim 1.5 \ {\rm GeV}$
it appears appropriate to 
treat charmed baryons at leading order 
as being composed of an infinitely 
massive charm quark surrounded by 
light degrees of freedom with mass of order 
$\Lambda_{\rm QCD}$.
Corrections to the infinite mass limit
can be computed or estimated in a systematic
expansion in $1/M_c$.
In the heavy quark limit the spin of the light degrees of
freedom becomes a good quantum number because the spin dependent 
interactions between the charm
quark and the light degrees of freedom are suppressed by 
inverse powers of $m_c$.
Hence, the charmed baryons can be classified by the spin of 
the light degrees of freedom,
$s_l$.
The baryons can be further classified by their transformation 
properties under the light
quark flavor symmetry SU(3).
The lowest lying baryons have the quantum numbers of 
two light quarks with no  orbital excitation
and consequently form two irreducible representations of 
flavor SU(3), 
the $\overline{\bf 3}$ and the ${\bf 6}$.
The baryons in the $\overline{\bf 3}$ must have $s_l=0$ 
while those in the 
${\bf 6}$ must have $s_l=1$ by the antisymmetry of the state.
Combining $s_l$ with the heavy quark spin produces baryons
in the ${\bf 6}$ with $J^\pi={3\over 2}^+$, denoted ${\bf 6^*}$,
and $J^\pi={1\over 2}^+$, as well as 
baryons in the $\overline{\bf 3}$ with $J^\pi={1\over 2}^+$.
In the infinite charm mass or light quark SU(3) limit of 
QCD the flavor eigenstates become the
mass eigenstates.
However, in the real world neither symmetry is exact and 
there is mixing between the two
multiplets.
Mixing between the $\Lambda_c^+$ member of the 
$\overline{\bf 3}$ and the 
$\Sigma_c^+$ member of the ${\bf 6}$ is strongly 
suppressed by isospin symmetry
and heavy quark symmetry and we will neglect it.
Also, the $\Omega_c^0$ member of the  ${\bf 6}$ does not 
have a corresponding state in the  $\overline{\bf 3}$ with 
which to mix.
However, the $\Xi_{c3}$ of the  $\overline{\bf 3}$ and the 
$\Xi_{c6}$ states of the ${\bf 6}$ can mix via SU(3) breaking 
and heavy quark spin symmetry
breaking interactions to form the mass eigenstates $\Xi_c$ 
and $\Xi_c^\prime$. The origin of this mixing and its implications
for $\Xi_b$ semileptonic decay is the subject of this paper.

In the context of the heavy quark effective theory,
the masses of the flavor eigenstates can be expanded
in powers of $1/M_c$ in terms of nonperturbative matrix elements
$\bar \Lambda, \lambda_1, etc.$. 
Mixing between the $\Xi_{c3}$ and $\Xi_{c6}$ breaks the heavy quark spin symmetry
and vanishes in the SU(3) 
symmetry limit. Since this mixing contributes to the masses
of the $\Xi_c$ and $\Xi^\prime_c$ at ${\cal O}\left( 1/M_c^2\right)$,
it appears unlikely its contribution can be determined from
a spectroscopic study of the charmed baryons
(this disagrees with the conclusion of reference \cite{ito}). 
In principle (though not in practice) 
the one $M_c^0$, two $1/M_c$ and six $1/M^2_c$ parameters in 
the $\Xi_c,\Xi^*_c$ baryon mass expansions can be extracted from 
inclusive semileptonic decay spectra\cite{incl} and substituted 
into the baryon mass expansions to yield the off-diagonal element
in the baryon mass-matrix.
Since some of these nonperturbative parameters are 
scheme-dependent quantities, 
we expect that naive definitions  of 
a mixing angle based on this off-diagonal element will be  
similarly scheme-dependent. For this reason it is important
to consider physical observables, such as semileptonic decays,
when discussing mixing in a field-theoretic context.

In this paper we compute the leading contributions to 
$\Xi_b\rightarrow \Xi_c^\prime l\overline{\nu}_l$
and $\Xi_b\rightarrow \Xi_c^{*} l\overline{\nu}_l$
in heavy hadron chiral perturbation theory. We contrast 
this with the nonrelativistic quark model, where these decays
arise entirely from mixing. It turns out that while
the physical mechanisms underlying both the quark model
and our calculation are the same, their implications for
observables can be quite different. 

In the non-relativistic quark model
the mixing between the $\Xi_{c6}$ and $\Xi_{c3}$ arises from 
the chromomagnetic hyperfine interaction
between the charmed quark and the light degrees of freedom
of the form
\begin{eqnarray}\label{QMint}
{\cal H}_{\rm int} & = & A\ 
\sum_i { {\bf s_c}\cdot {\bf s_j}\over M_C M_i}
\ \ \ ,
\end{eqnarray}
where $M_c $ and $M_i$ are the 
constituent quark masses and $A$ is a model dependent constant.
This gives rise to a mixing angle $\theta_c^{\rm QM}$ that 
relates the mass and flavor eigenstates
\begin{eqnarray}\label{QMwaves}
|\Xi_c\rangle & = & \cos\theta_c^{QM} |\Xi_{c3}\rangle
\ +\ \sin\theta_c^{QM}|\Xi_{c6}\rangle
\nonumber\\
|\Xi_c^\prime \rangle & = & -\sin\theta_c^{QM} |\Xi_{c3}\rangle
\ +\ \cos\theta_c^{QM}|\Xi_{c6}\rangle
\ \ \ .
\end{eqnarray}
This definition is universal in the sense that any
SU(3) and heavy quark spin symmetry violating
process involving these states 
resulting from a single insertion of a quark bilinear 
will depend only upon $\theta_c^{\rm QM}$.
For instance, radiative and semileptonic decay 
processes between these states that are forbidden 
in the heavy quark and flavor symmetry
limit depend on $\theta_c^{\rm QM}$.
However, in field theory, the effects of SU(3) and heavy quark spin 
symmetry breaking interactions
cannot be described by one such parameter as there are 
both wavefunction and vertex
modifications that need not be the same for all processes.

We consider the semileptonic decay 
$\Xi_b\rightarrow \Xi_c^\prime l \overline{\nu}_l$,
which vanishes in the heavy quark ($m_{b,c}\rightarrow \infty$)
and SU(3) limits of QCD. For our purposes it is sufficient to
examine only the zero recoil point.
This decay has the nice property that the matrix 
element of the vector current
vanishes when $m_b = m_c$ due to the orthogonality of the 
states at this symmetry
point (there is an exact global SU(2)$_H$ for the heavy 
quarks when $m_c=m_b$).
In the naive quark model one finds that the relevant matrix element is
\begin{eqnarray}\label{mixangledef}
\langle \Xi_c^\prime | 
\overline{c}\gamma_\mu (1-\gamma_5)b 
|\Xi_b\rangle & = & 
- \left(\sin\theta_c^{\rm QM} - \sin\theta_b^{\rm QM}\right)
\overline{U}_{\Xi_c}\gamma_\mu U_{\Xi_b}
\nonumber\\
& + &  
\left(\sin\theta_c^{\rm QM} + {1\over 3} 
\sin\theta_b^{\rm QM}\right)
\overline{U}_{\Xi_c}\gamma_\mu\gamma_5 U_{\Xi_b}
\ \ \ \  ,
\end{eqnarray}
where $\theta_b^{\rm QM} \sim  M_C/M_B\  \theta_c^{\rm QM}$ is 
the corresponding angle in the $b$-sector.
With the observation of the $\Sigma_c^*$ and a 
measurement of the mass difference
to the $\Sigma_c$ we have a measure of the size of 
spin-symmetry breaking
interactions in the charmed baryon sector.  
Fitting to a naive quark model  (such as used in \cite{ms})
gives a mixing angle of $\theta_c^{QM} \sim - 2^o$. 
A quark model calculation utilizing $SU(4)$ flavor 
symmetry\cite{franklin} yields an angle of 
$\theta_c^{QM} \sim - 4^o$.

Given the $\Sigma^*_c - \Sigma_c$ mass difference, we can 
study the effects of the chromomagnetic hyperfine
interaction in the model-independent framework of heavy
hadron chiral perturbation theory. 
In particular, the one operator at leading order in the chiral 
expansion that gives the $\Sigma^*_c - \Sigma_c$ mass difference will also
contribute to $\Xi_b\rightarrow 
\Xi_c^\prime l \overline{\nu}_l$ decay through one loop
graphs involving a $K, \eta$ or $\pi$. 
Other heavy quark symmetry violating effects, such as current and
coupling modifications, will also feed into
$\Xi_b\rightarrow
\Xi_c^\prime l \overline{\nu}_l$ decay through one loop
graphs, but we will see that they are formally sub-leading. 
While numerically the sub-leading contributions may be
comparable to the hyperfine contribution, one 
expects the contribution of the leading term to be 
indicative of the size of the actual decay rate.
Quark model expectations that the hyperfine interaction 
dominates over all other effects reinforce this idea. We
will see that neither expectation seems to be realized.

In the infinite mass limit the leading order 
interactions of the charmed baryons
with the pseudo-Goldstone bosons associated with 
chiral symmetry breaking are given by  
\cite{choa}
\begin{eqnarray}
{\cal L}^{(0)} & = & 
i\overline{T} v\cdot {\cal D} T
- i \overline{S}_\alpha  v\cdot {\cal D} S^\alpha
-\Delta_0 \overline{S}_\alpha  S^\alpha
\nonumber\\
& + &  i g_2 \varepsilon_{\mu\nu\alpha\beta} 
\overline{S}^\mu v^\nu A^\alpha S^\beta
\ +\  g_3 \left( \overline{T} A_\mu S^\mu\ +\ {\rm h.c.} \right) 
\ \ \ ,
\end{eqnarray}
where we are using  heavy baryon fields with four-velocity $v_\mu$.
The mass difference between the triplet $T$ and sextet $S$ baryons 
is $\Delta_0 \approx 210 {\rm MeV}$ and
the axial vector field of mesons, denoted by 
$A_\mu$, is 
\begin{eqnarray}
A_\mu & = & 
{i\over 2}
\left(   \xi \partial_\mu \xi^\dagger - \xi^\dagger \partial_\mu
\xi \right) 
\nonumber \\
\xi & = & \exp \left( i M / f  \right)
\ \ \  ,
\end{eqnarray}
where $M$ is the usual octet of pseudo-Goldstone bosons, and
$f \approx 132 {\rm MeV} $ is the pion decay constant.
The axial couplings $g_2$ and $g_3$ are not determined 
by the symmetries and must
be fit to data.   
The recent measurements of the $\Sigma_c^*$ width 
($\Gamma(\Sigma_c^{*++}) = 17.9{+3.8\atop -3.2}\pm 4.0\ {\rm MeV}$,   
$\Gamma(\Sigma_c^{*0}) = 13.0{+3.7\atop -3.0}\pm 4.0\ {\rm MeV}$ )
\cite{CLEOb}
determines that
$|g_3| = 0.9\pm 0.2$, approximately $30\%$ smaller than one 
would have estimated from
large $N_c$ arguments.
The above Lagrange density embodies the spin symmetry 
of the heavy quark limit of
QCD and therefore the baryons in the ${\bf 6}$ and 
${\bf 6}^*$ representations are
degenerate at this order.
The observed mass difference is described by the inclusion of 
the explicit spin-symmetry breaking term
\begin{eqnarray}
{\cal L}^{(1)} & = &  {i\over 3}\  \delta_c\   
\overline{S}^\mu \sigma_{\mu\nu} S^\nu
\ \ \ 
\end{eqnarray}
for the charmed baryons and an analogous term for the b-baryon.
This operator contains precisely the same physics as the quark
model operator Eq.~(\ref{QMint}), namely the chromomagnetic hyperfine
interaction, so by computing the contribution of ${\cal L}^{(1)}$
to $\Xi_b\rightarrow \Xi_c^\prime e\overline{\nu}$ we capture
the essential physics underlying the quark model in a field-theoretic
context.

Matrix elements of flavor-changing currents have a chiral representation
that in general involves unknown, nonperturbative Isgur-Wise form functions.
At zero recoil heavy 
quark symmetry normalizes these form factors to unity allowing
us to compute the contribution from ${\cal L}^{(1)}$ in terms
of $g_2, g_3$, and physical masses.

At leading order in chiral perturbation theory, the amplitude for 
$\Xi_b\rightarrow \Xi_c^\prime e\overline{\nu}$
receives contributions from the mixing graph of
fig. 1 (wavefunction renormalization), 
as well as the vertex modification in fig. 2.
The contribution from the wavefunction renormalization 
is process-independent, and can be identified with
the contribution from the mixing angle in the quark model.
However, the process-dependent vertex contribution, which 
is naively the same size as the contribution from wavefunction
renormalization, has no analogue in the quark model.
Therefore we expect the prediction of chiral perturbation theory
to differ substantially from that of quark models, even
if the hyperfine interaction captures all the essential physics
underlying the $\Xi_b\rightarrow
\Xi_c^\prime l \overline{\nu}_l$ decay (that is, even if the operators
${\cal L}^{(1)}$, ${\cal H}_{\rm int}$ dominate the rate).

In the formal chiral limit, where the chiral symmetry breaking
scale 
$\Lambda_{\chi}\sim \Delta_0 \gg M_K, M_\pi$,
we expect the one-loop contribution to the
$\Xi_b\rightarrow \Xi_c^\prime$ 
matrix element to be of order
$\sim g_2 g_3 \delta_c M_K^2\log (M_K^2/\Delta_0^2) 
/(\Lambda_\chi^2 \Delta_0) \sim 0.02$,
yielding a mixing angle of a few degrees. 
In this limit the contributions of $1/M_c$ current and
coupling modifications are easy to compute because they
involve the same loop integral as the hyperfine interaction,
except that the hyperfine mass splittings $\delta_c, \delta_b$
may be set to zero. Explicit calculation shows that these
integrals have no $ M_K^2 \log{M_K^2}$ chiral logs, so
they are sub-leading to the hyperfine contribution.
Experimentally $\Delta_0$ is of the same order as $M_K$, 
so we might instead consider
the limit, $\Delta_0\sim M_K \sim m_s^{1/2}$ (In 
the large $N_c$ limit of QCD, $\Delta_0 \rightarrow 0$, so we may 
imagine this as a combined large $N_c$ and chiral limit).
In this case,
we expect to find a matrix element of order
$g_2 g_3 \delta_c M_K/\Lambda_\chi^2$,
yielding a similarly sized angle. Based on dimensional
analysis, contributions from $1/M_c$ current or coupling
modifications will be higher order,
$\sim (\Lambda_{QCD}/M_c) (M_K^2 \log M_K^2/\Lambda_\chi^2)$,
so the hyperfine interaction is the leading contribution
in this limit as well.
Notice that in  both cases these 
estimates are larger
than what one would naively estimate from local counterterms, 
whose leading contribution is of order
$m_s/m_c$.

The graph in fig. 1 causes the $\Xi_{c3}$ and $\Xi_{c6}$ flavor
eigenstates to mix in such a way that a non-unitary change of
basis is required to diagonalize the Hamiltonian. Nevertheless,
the off-diagonal element,
\begin{eqnarray}
A_{36}  &=& -{ g_2 g_3 \delta_c\over 2\sqrt{6}\  16 \pi^2 f^2 \Delta_0}
 \sum_i \  C_i\ \left[ K (m_i,0,\delta_c) 
    - K(m_i,\Delta_0,\Delta_0+\delta_c)\right],
\end{eqnarray}
is a scheme-independent, universal contribution to
SU(3) and heavy quark spin symmetry violating processes involving
these states, so it is tempting to
identify it with the mixing angle of the quark model.
The sum is over the mesons in the loop, with
Clebsch-Gordan coefficients $C_\pi = 3$, $C_K = -2$ and $C_\eta = -1$.
The loop integral function that arises,
keeping only terms that survive the
summation, is
\begin{eqnarray}\label{Kdef}
K(m,\delta_1,\delta_2) & = &
{2\over 3}
\left[
(\delta_1^2+\delta_2^2+\delta_1\delta_2)\log {m^2\over \mu^2}
+ { (\delta_1^2-m^2)^{3/2} \over \delta_1-\delta_2}
\log\left({ \delta_1-\sqrt{\delta_1^2-m^2}\over
\delta_1+\sqrt{\delta_1^2-m^2}}\right)
\right.
\nonumber\\
& & \left.
- { (\delta_2^2-m^2)^{3/2} \over \delta_2-\delta_1}
\log\left({ \delta_2-\sqrt{\delta_2^2-m^2}\over
\delta_2+\sqrt{\delta_2^2-m^2}}\right)
\right]
\ \ \ ,
\end{eqnarray}
where $\mu$ is a subtraction scale that will cancel out after
summing over $SU(3)$ states.
For physical mass values, comparing $A_{36}$ with the naive quark model
$| \sin\theta_c^{\rm QM} |$ 
gives a mixing angle of $1.4 g_2 g_3 $ degrees, consistent with
quark model expectations.

However, the hyperfine interaction induces not only mixing,
but also the vertex modification of fig. 2.
Combining the two and evaluating 
at zero-recoil ($v\cdot v^\prime=0$), we find a 
$\Xi_b\rightarrow \Xi_c^\prime e\overline{\nu}$
amplitude of 
\begin{eqnarray}\label{theanswer}
& & \langle \Xi_c^\prime | 
\overline{c} \gamma_\mu (1-\gamma_5) b 
|\Xi_b\rangle  =  
{ g_2 g_3\over 2\sqrt{6}\  16 \pi^2 f^2 }
\left[
F_V\ \overline{U}_{\Xi_c} \gamma_\mu U_{\Xi_b}  
\  +  \ 
F_A\ \overline{U}_{\Xi_c} \gamma_\mu \gamma_5 U_{\Xi_b}
\right]
\nonumber\\
F_V & = & \sum_i \  C_i\ \left[
K (m_i,\delta_c,\Delta_0+\delta_b) - K (m_i,0,\Delta_0)
 + {\delta_c\over 2\Delta_0} 
\left( K (m_i,0,\delta_c) - K(m_i,\Delta_0,\Delta_0+\delta_c)\right)
\right.
\nonumber\\
& & \left.
+ {\delta_b\over 2\Delta_0} 
\left( K (m_i,0,\delta_b) - K(m_i,\Delta_0,\Delta_0+\delta_b)\right)
\right]
\nonumber\\
F_A & = & {1\over 2} \sum_i \  C_i\ \left[
{2\over 9} K(m_i,0,\Delta_0) 
+ {16\over 9} K(m_i,0,\Delta_0+\delta_b)
- {10\over 9} K (m_i,\delta_c,\Delta_0+\delta_b) 
- {8\over 9}K (m_i,\delta_c,\Delta_0)
\right.
\nonumber\\
& & \left. - {\delta_c\over \Delta_0} 
\left( K (m_i,0,\delta_c) - K(m_i,\Delta_0,\Delta_0+\delta_c)\right)
+ {\delta_b\over 3\Delta_0} 
\left( K (m_i,0,\delta_b) - K(m_i,\Delta_0,\Delta_0+\delta_b)\right)
\right]
\ \ \ ,
\end{eqnarray}

For physical values, the mixing and vertex contributions nearly
cancel, resulting in an amplitude of
\begin{eqnarray}
& & \langle \Xi_c^\prime |
\overline{c} \gamma_\mu (1-\gamma_5) b
|\Xi_b\rangle  \approx 
10^{-3} g_2 g_3 \left[ 0.9\  
\overline{U}_{\Xi_c} \gamma_\mu U_{\Xi_b}
\ +\   3.0 \ 
\overline{U}_{\Xi_c} \gamma_\mu \gamma_5 U_{\Xi_b}
\right]
\ \ \ \  .
\end{eqnarray}
Notice that the amplitude given in 
Eq.~(\ref{theanswer}) is not in the form expected by
Eq.~(\ref{mixangledef}). 
The relative contributions of the
four terms in the naive 
quark model result, Eq.~(\ref{mixangledef}), are a 
consequence of including
one insertion of the hyperfine interaction, while
Eq.~(\ref{theanswer}) includes the effects of the hyperfine
interaction to all orders.
Treating the hyperfine splittings $\delta_{c,b}$ as small 
compared to the other mass scales
in the problem and expanding to linear order
we can write the zero-recoil matrix  element in the
form of Eq. (\ref{mixangledef}) with an effective (but
non-universal) mixing angle of
\begin{eqnarray}
sin(\theta_c^{eff}) = -{ g_2 g_3\over 4\sqrt{6}\  16 \pi^2 f^2 }
\Delta_0 \delta_c \sum_i\ C_i\ G(m_i , \Delta_0),
\end{eqnarray}
where
\def\do{\Delta_0}
\begin{eqnarray}
\int\ {d^n q\over (2\pi)^n} 
{q^a q^b\over [q^2-m^2][v\cdot q]^2[v\cdot q-\Delta_0]^2}
& = & 
{-i\over 16\pi^2} \left[ 
G(m,\Delta_0) g^{ab} \ +\ 
V(m_i,\Delta_0) v^a v^b 
\right],
\nonumber\\
G(m,\Delta_0) = {2\over 3 \do^3} \left[
\sqrt{\do^2-m^2} (2 m^2 \right.  &+& \left. \do^2)
\ln{ \do + \sqrt{\do^2 -m^2} \over \do - \sqrt{\do^2 -m^2} }
+ \do^3 \ln{m^2\over \mu^2} +2 m^3 \pi - 4 \do m^2 \right].
\ \ \ 
\end{eqnarray}
Numerically, this gives $ \theta_c^{eff} = -0.06 g_2 g_3$ degrees.

We can also compute the contribution of the hyperfine
interaction to $\Xi_b\rightarrow \Xi_c^* l \overline{\nu}_l$,
\begin{eqnarray}
& & \langle \Xi_c^* |
\overline{c} \gamma_\mu (1-\gamma_5) b
|\Xi_b\rangle  =
{2\over 9}{ g_2 g_3\over 16 \pi^2 f^2 }
\sum_i
\left[
K(m_i, \Delta_0 +\delta_b, \delta_c) -K(m_i, \Delta_0 , \delta_c)
\right. 
\nonumber\\
&+& \left. 5 K(m_i, \Delta_0 +\delta_b, 0) -5 K(m_i, \Delta_0 , 0)
+ {3 \delta_b \over \Delta_0}\left[ K(m_i, \delta_b,0)
         - K(m_i,\Delta_0, \Delta_0 + \delta_b) \right]
\right] \overline T^\mu_{\Xi_c} U_{\Xi_b} ,
\end{eqnarray}
where $T^\mu_{\Xi_c}$ is the Rarita-Schwinger spinor for the
$\Xi_c^*$.  Notice that this amplitude vanishes when
$\delta_b=0$, as required by angular momentum conservation.
For physical values, the mixing and vertex
contributions again nearly cancel, giving an amplitude of 
\begin{eqnarray}
\langle \Xi_c^* |
\overline{c} \gamma_\mu (1-\gamma_5) b
|\Xi_b\rangle  & = &
5 \cdot 10^{-4}\  g_2 g_3 \quad \overline T^\mu_{\Xi_c} U_{\Xi_b} 
\ \ \ ,
\end{eqnarray} 
again much smaller than naive quark model estimates.

The amplitudes arising from this calculation are roughly
a factor of thirty smaller than naive expectations.
In the limit 
$\do >> m_i, \delta_c$, the decay amplitude vanishes like
$\delta_c/\do$, since the intermediate loop baryon becomes
very massive, but we know of no a priori reason why the
amplitude would vanish for $\do \sim m_i$.
Algebraically, the strong cancelations between
mixing and vertex contributions arise because 
$K(m, \delta_1, \delta_2)$ in Eq.~(\ref{Kdef}) is 
a slowly varying function. Away from zero recoil, the 
vertex and mixing contributions get multiplied by different
nonperturbative form factors, so far from this kinematic point,
the contribution from
the hyperfine interaction could be an order of magnitude larger.

What can we conclude from the exceptionally small amplitudes
at zero recoil?
The formally leading contributions we have computed are 
clearly not the numerically dominant
terms: Other operators, responsible for current and 
coupling corrections, determine the rate. The ``estimate" 
provided by computing the hyperfine contribution cannot be trusted,
even to an order of magnitude. Nevertheless, we can learn something
interesting from the computation.

Firstly, naive power counting in the heavy baryon sector can
fail rather spectacularly. This can be traced in part to the
presence of the additional mass scale $\do$, which 
leads to powers of ${m_\pi \over \do}$ and ${\do \over m_K}$
that are important numerically. In addition, it is important
to keep powers of ${\delta_c \over m_K}$ (as we did in this
calculation), because this ratio
is neither parametrically small nor large in the most general
combined heavy quark, chiral limits. This feature has been
noted previously in the context of heavy mesons\cite{hmesons}.
Care should be used when
using chiral perturbation theory to compute properties
of charmed and bottom baryons ( e.g. \cite{mscpt}). 

Secondly, this computation reveals potential  drawbacks
to the naive quark model estimate. 
The quark model asserts that the
chromomagnetic hyperfine interaction is responsible for
$\bar 3 -6$ mixing, and that this mixing is solely responsible
for $\Xi_b\rightarrow \Xi_c^\prime l\overline{\nu}_l$. 
We have computed the contribution of the hyperfine interaction
to this decay. We find a process and scheme-independent
contribution from wavefunction mixing 
that can be identified with the quark model
mixing angle, and is numerically comparable to quark 
model expectations. 
However, the hyperfine interaction also induces
a compensating vertex modification to 
$\Xi_b\rightarrow \Xi_c^\prime l\overline{\nu}_l$
that is absent in the quark model, and nearly cancels 
the contribution from mixing. This suggests
the hyperfine interaction plays an inessential role
for this decay, at least near zero recoil. 
The decay may simply be dominated by operators that are not
addressable within the context of a nonrelativistic quark model.
 
These cautionary notes for both quark model and chiral perturbation
theory calculations highlight the difficulty of reliably
estimating the  $\Xi_b\rightarrow
\Xi_c^\prime l \overline{\nu}_l$ decay rate. By illustrating
the limitations inherent in both methods, we may aid
their successful application elsewhere. It will be interesting
to see if the remarkable cancelations plaguing the
contributions of the leading operator to $\Xi_b\rightarrow
\Xi_c^\prime l \overline{\nu}_l$ and $\Xi_b\rightarrow
\Xi_c^{*} l \overline{\nu}_l$ decay amplitudes are 
mirrored in other physical observables in the charmed baryon system.

\vskip 1.5cm

\centerline{\bf Acknowledgments}

We thank M. Wise and J. Walden for useful discussions.
This work is supported by the Department of Energy under Grant
no. DE-FG02-91-ER 40682. 
CGB  and ML  thank the nuclear theory group 
at the University of Washington for their hospitality during
part of this work.

\begin{figure}
\epsfxsize=8cm
\hfil\epsfbox{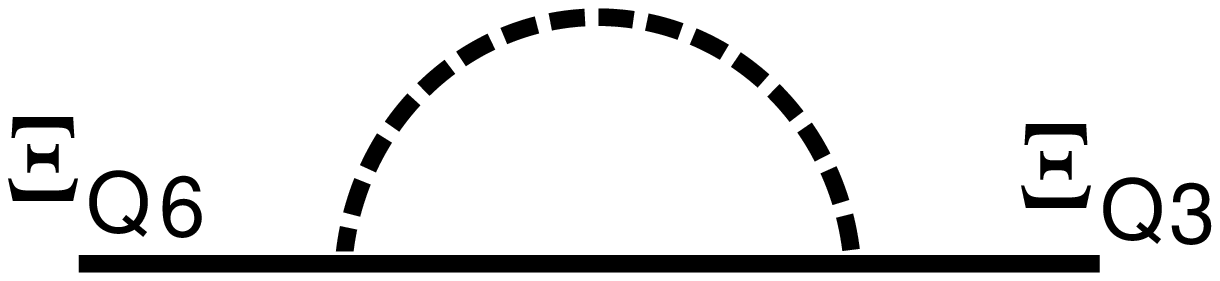}\hfill
\caption{ Wavefunction mixing between the charmed baryon or b-baryon states.
The dashed line denotes a pseudo-Goldstone boson.
The spin-symmetry violating mass difference between the intermediate states of the 
${\bf 6}^{(*)}$ representations prevents these graphs from vanishing.}
\label{wavefig}
\end{figure}

\begin{figure}
\epsfxsize=8cm
\hfil\epsfbox{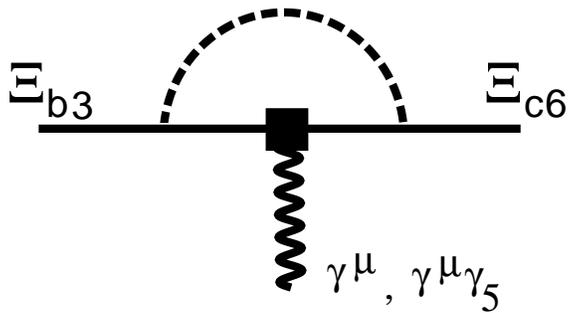}\hfill
\caption{ Vertex contribution to the decay $\Xi_b\rightarrow\Xi_c^\prime
e^-\overline{\nu}_e$. 
The dashed line denotes a pseudo-Goldstone boson. }
\label{wavefig}
\end{figure}

\end{document}